\documentstyle[aps,epsf]{revtex}
\begin{document}
\draft
\title{Linked Cluster Expansion Around Mean-Field Theories of Interacting
Electrons}
\author{V.~Jani\v{s}}
\address{Institute of Physics, Academy of Sciences of the Czech Republic,\\
CZ-18040 Praha 8,  Czech Republic}
\author{J.~Schlipf}
\address{Institut f\"ur Theoretische Physik C, Technische Hochschule
Aachen, \\
D-52056 Aachen, Federal Republic of Germany}
\maketitle

\begin{abstract}
A general expansion scheme based on the concept of linked cluster
expansion from the theory of classical spin systems is constructed for
models of interacting electrons. It is shown that with a suitable
variational formulation of mean-field theories at weak (Hartree-Fock) and
strong (Hubbard-III) coupling the expansion represents a universal and
comprehensive tool for systematic improvements of static mean-field
theories. As an example of the general formalism we investigate in detail an
analytically tractable series of ring diagrams that correctly capture
dynamical fluctuations at weak coupling. We introduce renormalizations of
the diagrammatic expansion at various levels and show how the resultant
theories are related to other approximations of similar origin. We
demonstrate that only fully self-consistent approximations produce global
and thermodynamically consistent extensions of static mean field theories. A
fully self-consistent theory for the ring diagrams is reached by summing the
so-called noncrossing diagrams.

\medskip
Preprint-no. RWTH-ITP-C 5/95
\end{abstract}

\section{Introduction}

A first step towards a global description of a model that cannot be
solved exactly is a mean-field theory (MFT). It usually shares many features
with an exact solution, offers a global phase diagram, and enables to
investigate the existence of various solutions that exhibit long-range
order. A comprehensive MFT represents a conserving and thermodynamically
consistent approximation, i.e.\ there exists an explicit free energy
functional applicable in the entire range of input parameters.

In the theory of interacting electrons it was for a long time only
the weak-coupling Hartree-Fock solution that fulfilled the usual
requirements imposed on a MFT in classical statistical mechanics of lattice
spin systems \cite{vjdv93}. Attempts to construct a strong-coupling MFT
around the atomic limit with analogous comprehensive properties failed in
most cases to produce a thermodynamically consistent MFT. In the case of the
Hubbard model mean-field theories reproducing the atomic limit have a
structure of the Hubbard-III solution \cite{H-III} and are commonly called
alloy-analogy approximations \cite{mor85,czy86}. These approximations suffer
from the deficiency that they do not represent a conserving approximation in
the sense of Baym \cite{ba62}. Only lately in connection with the limit of
high spatial dimensions, $d\rightarrow \infty $, a thermodynamically
consistent extension of the Hubbard-III solution with a free-energy
functional was proposed \cite{vjdv92}.

However, having a thermodynamically consistent MFT alone is
insufficient to make the mean-field results reliable. Unless we have an
expansion scheme around the MFT with which we can study the stability of the
mean-field solution, we cannot really judge whether and in which limit the
mean field produces physically relevant results. Hence it is part of the
construction of a comprehensive MFT to find an appropriate perturbation
theory the first term of which is just the mean-field grand potential. It is
fairly known that the Hartree-Fock MFT is a first, self-consistent
approximation of the weak-coupling perturbation expansion. It is, however,
much more difficult to find an expansion scheme the first step of which
would be a MFT of the Hubbard-III type, since this MFT cannot be mapped onto
a Fermi gas. We cannot use a usual expansion relying on Gaussian integration
and the Wick theorem.

We are hence facing the problem how to connect a general MFT with an
appropriate perturbation theory. First attempts to abandon the weak-coupling
expansion in the Hubbard model were made by Hubbard \cite{hu66} who
formulated an unrenormalized expansion in the hopping amplitude within which
he rederived an approximation now called Hubbard-I. Recently Metzner \cite
{metz91} using a formulation in $d=\infty $ extended the Hubbard approach
and showed a way how renormalizations from the linked cluster expansion \cite
{wor74} can be introduced. However, it has proved cumbersome to deal with
renormalizations effectively because of the quantum character of the model.
This expansion does not efficiently go beyond the Hubbard-III solution and
in this form it seems inadequate to improve strong-coupling mean-field
theories.

It is the aim of this paper to formulate an expansion scheme
suitable to encompass general mean-field theories, the Hartree-Fock at weak
coupling as well as the Hubbard-III at strong coupling, which enables to
introduce consistently renormalizations leading to conserving approximations
in the Baym sense. Importance of a systematic
expansion around a comprehensive MFT was recently acknowledged in \cite
{vjdv93}, where a formal perturbation expansion resulted from a specific
variational formulation of mean-field theories being either an upper bound
(Hartree-Fock) or a lower bound (extension of Hubbard-III from \cite{vjdv92}%
). In the following we formulate this expansion explicitly, show how to
generate various approximations, how to introduce renormalizations, and
study quantitatively the effects of dynamical fluctuations on the results of
static mean-field theories. We use the linked cluster expansion (LCE) from
the classical statistical theory of liquids and lattice spin systems
reviewed in \cite{wor74}. The LCE is applied here in quite a different
manner than in \cite{metz91}, since we cannot expand in the hopping
amplitude. We formulate the mean-field theories in such a way that we gain a
universal expansion scheme for the weak- and strong-coupling theories.
Within the derived scheme we compare different approximations having closed
form. We further discuss the role and importance of self-consistencies in
the light of the recent claim that a partial self-consistency, called
iterated perturbation theory, may be superior to the fully self-consistent
approach at intermediate and strong coupling \cite{gk92}. We conclude that
only fully self-consistent theories represent conserving approximations.
As the main result we then
derive a fully self-consistent version of the ring-diagram approximation
by summing the so-called ``noncrossing'' diagrams for the grand potential
and thereby win a global, thermodynamically consistent theory containing
dynamical fluctuations that is equally applicable to weak as well to strong
coupling MFT's. In this way we remove the doubts that renormalized RPA or
more advanced self-consistent approximations may not lead to conserving
approximations \cite{ls69,ha69}.

The paper is organized as follows. In Sec.\ II we reformulate the
variational approach to the construction of the comprehensive MFT from \cite
{vjdv93} so that the same expansion scheme can be used for the sample
mean-field theories of interacting electrons: Hartree-Fock and the
thermodynamically consistent extension of Hubbard-III from \cite{vjdv92}.
Sec.\ III summarizes the formalism of the LCE needed to generate the
appropriate expansion for the grand potential of itinerant models with local
electron-electron interaction. In Sec.\ IV we choose a special class of
diagrams, ring diagrams, leading to an approximation in closed form and
being exact up to second order in the interaction strength. In Secs.\ V and
VI we derive one-particle properties from the variational grand potential
and present numerical results obtained from various simplifications of the
ring-diagram approximation. Here the position of the iterated perturbation
theory from \cite{gk92} in the present scheme is thoroughly discussed. In
Sec.\ VII a controlled way to reach full self-consistency and thereby a
thermodynamically consistent and conserving theory of the ring-diagram
approximation is presented.

\section{Variational formulation of mean-field theories}

A MFT is controllable only if it enables a systematic expansion
towards the full solution. Not all constructions of mean-field like
approximations are suitable for further corrections via perturbation
expansions. E.g.\ heuristic derivations of the Gutzwiller extension of the
Hartree-Fock MFT \cite{gu64} or of a strong-coupling MFT \cite{H-III} do not
seem to fit into a systematic expansion scheme starting with the MFT. On the
other hand, a variational formulation of mean-field theories from \cite
{vjdv93} offers a consistent way to connect a MFT with a perturbation
expansion. We present in this section an analogous variational formulation
of mean-field theories of itinerant electrons that enables to formulate a
universal expansion scheme for both weak and strong coupling. Explicitly we
will refer to the weak-coupling (Hartree-Fock) and strong-coupling (Ref.
\cite{vjdv93}) theories of the Hubbard model.

For an expansion around a MFT we must decompose the total
Hamiltonian into its unperturbed part (leading to the desired MFT) and a
perturbation
\begin{equation}
\label{eq1}H=H_0+\Delta H.
\end{equation}
Our aim is to find suitable constituents of the decomposition (\ref{eq1})
for the weak and strong coupling mean-field theories leading to conceptually
the same perturbation expansion.

We start with the simpler case of the Hartree-Fock approximation.
For the sake of simplicity we restrict ourselves to spatially homogeneous
solutions only. The variational parameters in the Hartree-Fock theory are
the particle densities $n_\sigma $ and we can write
\begin{mathletters}
\label{eq2}
\begin{equation}
\label{eq2a}
H=H_{HF}+\Delta H_{HF}
\end{equation}
with%
\begin{equation}
\label{eq2b}
H_{HF}=-{\cal N}Un_\uparrow n_\downarrow+\sum_{{\bf k},\sigma }\left(
\epsilon _{\bf k}+Un_{-\sigma }\right) a_{{\bf k}\sigma }^{\dagger }a_{{\bf
k}\sigma },
\end{equation}
\begin{equation}
\label{eq2c}
\Delta H_{HF}=U\sum_i\left( \widehat{n}_{i\uparrow
}-n_{\uparrow }\right) \left( \widehat{n}_{i\downarrow }-n_{\downarrow
}\right) ,
\end{equation}
\end{mathletters}
where $\widehat{n}_{i\sigma }=c_{{\bf i}\sigma }^{\dagger }c_{{\bf i}\sigma }
$ and ${\cal N}$ is the number of lattice sites. This decomposition leads to
the standard perturbation expansion around the Hartree-Fock solution
generated by the Hamiltonian $H_{HF}$.

The situation in the case of the strong-coupling MFT is more
complicated. The decomposition used in \cite{vjdv93} leading to a lower
bound does not have the structure of (\ref{eq1}). To obtain a structure
similar to (\ref{eq2}) we have to distinguish the static and dynamic degrees
of freedom used in the strong-coupling MFT. We decorate the static degrees
of freedom with an index $s$. We choose a decomposition of the Hubbard
Hamiltonian into two Falicov-Kimball subhamiltonians with equal weight $%
\lambda =1/2$ \cite{vjdv93}. We hence define
\begin{mathletters}
\label{eq3}
\begin{equation}
\label{eq3a}
H_\sigma =\sum_{{\bf k}}\left( \epsilon _{{\bf k}}+E_\sigma \right) a_{{\bf k}
\sigma }^{\dagger }a_{{\bf k}\sigma }-E_{-\sigma}
\sum_i\widehat{n}_{i-\sigma }^s +\frac U2\sum_i\widehat{n}_{i\sigma }%
\widehat{n}_{i-\sigma }^s \,,
\end{equation}
where $\widehat{n}_{i\sigma }^s =c_{{\bf i}\sigma }^{s\dagger }%
c_{{\bf i}\sigma }^s$ and
\begin{equation}
\label{eq3b}
\Delta H=\sum_{i,\sigma }E_\sigma \left( \widehat{n}%
_{i\sigma }^s -\widehat{n}_{i\sigma }\right) +\frac U2\sum_{i,\sigma }%
\widehat{n}_{i\sigma }\left( \widehat{n}_{i-\sigma }-\widehat{n}_{i-\sigma}^s
\right)
\end{equation}
from which the desired decomposition of the Hubbard Hamiltonian
directly follows:
\begin{equation}
\label{eq3c}
H=\sum_\sigma H_\sigma +\Delta H%
\,.\quad
\end{equation}
\end{mathletters}
It is easy to show that $F=-\beta ^{-1}\ln \mbox{Tr}\exp \{-\beta
\sum_\sigma H_\sigma \}$ in $d=\infty $ generates the mean-field theory of
\cite{vjdv93} with $\lambda _{at}=0,\lambda _\sigma =1/2$. The variational
parameters here are not $\widehat{n}_{i\sigma }^s$ (which are operators),
but energies $E_\sigma $. It is also clear from (\ref{eq3a}) that the
dynamics of $H_{\uparrow }$ and $H_{\downarrow }$ decouple and $%
F_{MF}=F_{\uparrow }+F_{\downarrow }$. The stationarity conditions for such
a MFT lead to
\begin{equation}
\label{eq4}\left\langle \widehat{n}_{i\sigma }\right\rangle
_{MF}=\left\langle \widehat{n}_{i\sigma }^s\right\rangle _{MF}\,.
\end{equation}
We have two different contributions to the perturbation $\Delta H\,$ in (\ref
{eq3b}), one quadratic and one quartic in creation and annihilation
operators. This makes perturbation expansion cumbersome. We can, however,
drop the quadratic term without changing the physics of the full Hubbard
Hamiltonian. The quadratic term only shifts the spin-dependent chemical
potential $\mu _\sigma =\mu +\sigma h$ of the mobile (dynamic) electrons
with the energy $E_\sigma $ playing the role of the chemical potential for
the local (static) electrons. These auxiliary local electrons completely
decouple from the physical, itinerant electrons in the full solution. The
energies $E_\sigma $ in the physical sector then redefine the origins for
the chemical potential $\mu $ and the magnetic field $h$. The physical
results do not differ from those obtained from the original Hubbard
Hamiltonian without energies $E_\sigma $. We can hence choose
\begin{mathletters}
\label{eq5}
\begin{equation}
\label{eq5a}
H=H_{MF}+\Delta H_{MF}=\sum_\sigma H_\sigma +\Delta H_{MF}
\end{equation}
and
\begin{equation}
\label{eq5b}
\Delta H_{MF}=\frac U2\sum_{i,\sigma}\widehat{n}_{i\sigma
}\left( \widehat{n}_{i-\sigma }-\widehat{n}_{i-\sigma }^s \right) \,.
\end{equation}
\end{mathletters}
It is worth noting that the decomposition (\ref{eq3})-(\ref{eq5}) enabling a
systematic perturbation expansion is not derivable for the MFT with $\lambda
_\sigma =1$ and $\lambda _{at}=-1$ as used in \cite{vjdv92,vjjmdv93}. To
apply decomposition (\ref{eq1}) consistently, no negative terms are allowed.
If we choose $\lambda_{at}=0$, the resultant MFT does not reproduce the
atomic limit exactly. Although the atomic limit is not covered by (\ref{eq5}),
due to the halving of the interaction strength in the subhamiltonians $%
\widehat{H}_\sigma$, (\ref{eq5}) still contains the split-band limit.
Important features of (\ref{eq5}) are that the mean-field free energy is an
exact lower bound to the free energy of the Hubbard model and we can expand
consistently around the MFT. Moreover, it was shown earlier in \cite{vjdv93}
that the choice $\lambda _\sigma =1$ and $\lambda _{at}=-1$ is equivalent at
$T=0$ to $\lambda _\sigma =1/2,\lambda _{at}=0$ with doubled interaction
strength $U$.

Both the mean-field theories for weak and strong coupling now have
formally identical structure and the corresponding perturbation expansion
can be formulated in the same manner. We use the LCE well elaborated in the
classical statistical mechanics \cite{wor74} as an appropriate tool for our
purposes.

\section{General formalism of the Linked Cluster Expansion}

The aim of the LCE is to represent the free energy (grand potential)
as an appropriate series generated by derivatives w.r.t.\ auxiliary external
sources added to the unperturbed Hamiltonian. Contrary to the standard
weak-coupling expansion, LCE does not rely on the Wick theorem and is hence
appropriate for the construction of a systematic expansion around any MFT.
The only restriction on applicability of the LCE is that the generating
unperturbed functional (free energy of the unperturbed system with suitable
external fields) has to be explicitly known.

Our aim is to formulate a perturbation expansion around quantum
mean-field theories using decompositions (\ref{eq2}) and (\ref{eq3}).
Although the unperturbed mean-field functional to (\ref{eq2}) is known in
any dimension, we restrict the expansion to infinite dimensions, since only
there the unperturbed functional to the decomposition (\ref{eq3}) can
explicitly be constructed. This restriction leads to a substantial
simplification of the perturbation expansion, since the effective electron
propagator in perturbation theory does not depend on momentum explicitly
\cite{vo93}.

The generating functional for a quantum LCE can be represented as
\begin{equation}
\label{eq7}g_{MF}^{(0)}\{\mu _\sigma ,E_\sigma \}=\frac 1{{\cal N}}\ln
\mbox{Tr} {\cal T}\exp \left\{ -\int\limits_0^\beta d\tau \left[ H_{MF}(\tau
)-\sum_{i,\sigma }\mu _\sigma (\tau )\widehat{n}_{i\sigma }(\tau )\right]
\right\}
\end{equation}
with ${\cal T}$ standing for the time ordering operator and Tr denoting the
trace in Fock space. Functional (\ref{eq7}) is related to the grand
potential $\Omega _{MF}^{(0)}=-\beta ^{-1}g_{MF}^{(0)}$. The generating
external sources were absorbed into the mean-field variational parameters, $%
n_\sigma$ for the weak coupling MFT and $E_\sigma$ and the chemical
potential of the itinerant electrons $\mu _\sigma $ for the strong-coupling
MFT. These parameters now become time dependent. They are coupled to the
time-dependent particle densities which are the only operators appearing in
the perturbation $\Delta H\,$.

The Hamiltonian in (\ref{eq7}) can explicitly be written for the
weak-coupling MFT as
\begin{mathletters}
\label{eq6}
\begin{equation}
\label{eq6a}
H_{HF}(\tau)=\sum_{{\bf k,}\sigma }\epsilon _{{\bf k}}a_{{\bf k}\sigma }^{
\dagger }(\tau)a_{{\bf k}\sigma }(\tau)+U\sum_{i,\sigma }%
n_{-\sigma }(\tau)\left( \widehat{n}_{i\sigma }(\tau)-n_\sigma \right)
+U{\cal N}n_\uparrow n_\downarrow \; ,
\end{equation}
and for the strong-coupling MFT as
\begin{equation}
\label{eq6b}
H_{MF}(\tau)=\sum_{{\bf k},\sigma }\epsilon _{{\bf k}}a_{{\bf k}\sigma }^{
\dagger }(\tau)a_{{\bf k}\sigma }(\tau)+\sum_{i,\sigma }
E_\sigma (\tau)\left( \widehat{n}_{i\sigma }(\tau)-\widehat{n}_{i\sigma }^s
(\tau)\right) +\frac U2\sum_{i,\sigma }\widehat{n}_{i\sigma }(\tau)
\widehat{n}_{i-\sigma }^s (\tau)\,.
\end{equation}
\end{mathletters}
The time dependency was added to the variational parameters only where
necessary. In the limit $d=\infty $ we obtain for the Hartree-Fock MFT
\begin{mathletters}
\label{eq8}
\begin{equation}
\label{eq8a}
g_{HF}^{(0)}\{ n_\sigma \}=\int\limits_0^\beta d\tau
\left\{ Un_{\uparrow }n_{\downarrow }+\sum_\sigma
\int\limits_{-\infty }^\infty d\epsilon \rho (\epsilon )\left[ \ln (\partial
/\partial \tau +\mu _\sigma -Un_{-\sigma }-\epsilon)\right]
(\tau ,\tau ^{+})\right\}\, ,
\end{equation}
while for the strong-coupling MFT the generating functional reads
\begin{eqnarray}
\label{eq8b}
&&g_{MF}^{(0)}\{\mu_\sigma ,E_\sigma \}=\sum_\sigma \int\limits_0^\beta d\tau
\left\{ \int\limits_{-\infty }^\infty d\epsilon \rho (\epsilon )\left[ \ln (
\partial /\partial \tau +\mu_\sigma-E_\sigma-\Sigma _\sigma -\epsilon
)\right](\tau ,\tau ^{+})\right. \nonumber\\
&&\left. +\left[ \ln G_\sigma \right](\tau ,\tau ^{+})
+E_\sigma (\tau)n_\sigma ^s (\tau)+(1-n_{-\sigma}^s (\tau))\left( \left[
\ln (\gamma_\sigma )\right] (\tau ,\tau ^{+})-
\beta ^{-1}\ln (1-n_{-\sigma }^s (\tau))\right)\right. \nonumber \\
&&\left. +n_{-\sigma }^s (\tau)\left( \left[ \ln \left(\gamma_\sigma
-\frac U2 \right) \right] (\tau ,\tau ^{+})-\beta ^{-1}
\ln n_{-\sigma }^s (\tau)\right) \right\}\; ,
\end{eqnarray}
\end{mathletters}
where $n_\sigma (\tau ,\tau ^{\prime }):=\delta (\tau -\tau ^{\prime
})n_\sigma (\tau )$, $E_\sigma (\tau ,\tau ^{\prime }):=\delta (\tau -\tau
^{\prime })E_\sigma (\tau )$, $\mu _\sigma (\tau ,\tau ^{\prime }):=\delta
(\tau -\tau ^{\prime })\mu _\sigma (\tau )$ and $\gamma _\sigma (\tau ,\tau
^{\prime }):=G_\sigma ^{-1}(\tau ,\tau ^{\prime })+\Sigma _\sigma (\tau ,\tau
^{\prime })$. The single-particle Green function $G_\sigma (\tau ,\tau
^{\prime })=-\langle {\cal T} c_\sigma (\tau )c_\sigma ^{\dagger }(\tau
^{\prime})\rangle _T$ and the corresponding self-energy $\Sigma _\sigma
(\tau ,\tau^{\prime })$ were used. With these expressions we can derive the
explicit LCE with the perturbation $\Delta H(\tau )$ if we use
\begin{mathletters}
\label{eq9}
\begin{equation}
\label{eq9a}
\Delta H_{HF}(\tau )=\frac 1U \frac \delta {\delta
n_{\downarrow }(\tau )}\frac \delta {\delta n_{\uparrow }(\tau )}
\,,
\end{equation}
\begin{equation}
\label{eq9b}
\Delta H_{MF}(\tau )=-\frac U2\sum_\sigma \frac \delta
{\delta \mu _\sigma (\tau )}\frac \delta {\delta E_{-\sigma }(\tau )}\,
\end{equation}\end{mathletters}
and represent the full potential in $d=\infty $ for the Hartree-Fock MFT as
\begin{mathletters}
\label{eq10}
\begin{equation}
\label{eq10a}
g_{HF}=\ln \left[ \exp \left\{ -\frac 1U\int\limits_0^\beta d\tau
 \frac \delta {\delta n_{\downarrow }(\tau )}\frac \delta {\delta
n_{\uparrow }(\tau )} \right\} \exp g_{HF}^{(0)}\{ n_\sigma \}\right]
\end{equation}
and for the strong-coupling MFT as
\begin{equation}
\label{eq10b}
g_{MF}=\ln \left[ \exp \left\{ \frac U2\sum_\sigma
\int\limits_0^\beta d\tau \frac \delta {\delta \mu _\sigma (\tau )}\frac
\delta {\delta E_{-\sigma }(\tau )}\right\} \exp g_{MF}^{(0)}\{ \mu _\sigma ,
E_\sigma \}\right] \,.
\end{equation}
\end{mathletters}
Note that the imaginary time is removed from the variational variables in
the end of all calculations to restore the time homogeneity.

The aim of the LCE is to define a direct expansion for the grand
potential, i.e. to interchange the logarithm with the variational
derivatives in formulas (\ref{eq10}). Since the derivatives are of second
order, this is a tremendous task and leads to sophisticated rules for the
construction of exact contributions in all orders of the perturbation $%
\Delta H$ \cite{wor74}. We can, however, formally write down the full
potential as
\begin{equation}
\label{eq11}g_{MF}=\exp \left\{ \frac U2\sum_\sigma \int\limits_0^\beta
d\tau \left[ \frac \delta {\delta \mu _\sigma (\tau )}\frac \delta {\delta
E_{-\sigma }(\tau )}\right] \right\} g_{MF}^{(0)}\{\mu _\sigma ,E_\sigma
\}\,.
\end{equation}
In (\ref{eq11}) the brackets with variational derivatives denote a linear
operator which commutes with the derivatives and acts on the unperturbed
functional $g_{MF}^{(0)}$ as
\begin{equation}
\label{eq12}\left[ \frac \delta {\delta \mu _\sigma (\tau )}\frac \delta {%
\delta E_{-\sigma }(\tau )}\right] g^{(0)}_{MF}\{\mu _\sigma ,E_\sigma \}:=%
\frac{\delta ^2g^{(0)}_{MF}\{\mu _\sigma ,E_\sigma \}\,}{\delta \mu _\sigma
(\tau )\delta E_{-\sigma }(\tau )}+\frac{\delta g^{(0)}_{MF}\{\mu _\sigma
,E_\sigma \}\,}{\delta \mu _\sigma (\tau )}\frac{\delta g^{(0)}_{MF}\{\mu
_\sigma ,E_\sigma \}\,}{\delta E_{-\sigma }(\tau )}\,.
\end{equation}

Definition (\ref{eq12}) expresses a derivative w.r.t. a bond
(nonlocal) variable (left-hand side) in terms of site (local) variables
genuine to the unperturbed functional $g^{(0)}_{MF}$ (right-hand side). The
power expansion of the exponential in (\ref{eq11}) together with the
definition of the bracketed derivative (\ref{eq12}) determines the
unrenormalized LCE. To save the space we make explicit derivations for the
strong-coupling MFT only. We hence drop the subscript ``MF'' in the
forthcoming formulas.

It is convenient as in other perturbation expansions to define a
graphical representation for the LCE and to deal with diagrams by
introducing vertices and internal lines. If we denote $x_\sigma ^\alpha
(\tau )$, $\alpha =\pm $ with $x_\sigma ^{+}(\tau )=\mu _\sigma (\tau )-\mu
_\sigma $ and $x_\sigma ^{-}(\tau )=E_\sigma -E_\sigma (\tau )$ we can
define an unrenormalized $n$-vertex function
\begin{equation}
\label{eq13}\Gamma _{(0)\,\sigma _1\cdots \sigma _n}^{(n)\,\alpha _1\cdots
\alpha _n}(\tau _1,\ldots ,\tau _n):=\frac{\delta ^ng^{(0)}\{\mu _\sigma
,E_\sigma \}}{\delta x_{\sigma _1}^{\alpha _1}(\tau _1)\cdots \delta
x_{\sigma _n}^{\alpha _n}(\tau _n)}\,
\end{equation}
and an unrenormalized internal line (edge) $v_{\sigma _1\sigma _2}^{\alpha
_1\alpha _2}(\tau _1,\tau _2)$ connecting the vertices. The ``propagator''
is hence a matrix in the internal degrees of freedom of the vertices, $\alpha
$, $\sigma $ and $\tau $. We can assign a graphical representation to both
the vertices and the edges. We denote an $n$-vertex as
\begin{mathletters}
\label{eq14}
\begin{equation}
\label{eq14a}
\begin{minipage}{2.5cm}
\begin{picture}(68.4,68.4)
\put(34.2,34.2){\circle{20}}
\put(29.0,30.5){$\Gamma _0$}
\put(38.0,43.5){\line(1,2){8}}
\put(51.2,56.1){1}
\put(44.2,34.2){\line(1,0){16}}
\put(62.0,25.5){2}
\put(27.0,27.0){\line(-1,-1){12}}
\put(18.5,7.3){n}
\put(53.9,17.1){\circle*{2}}
\put(45.4,11.5){\circle*{2}}
\put(34.5,9){\circle*{2}}
\end{picture}
\end{minipage}
\end{equation}
where $1=(\alpha _1,\sigma _1,\tau _1)$, \ldots \ . In this
version of the LCE the electron-electron interaction plays the role of
edges connecting the vertices.
We assign a dashed line to an edge, i.e.
\begin{equation}
\label{eq14b}
\begin{minipage}{7.3pt}
\begin{picture}(7.2,28.5)
\put(9.1,13){\circle{4}}
\put(6.5,0){1}
\end{picture}
\end{minipage}
------
\begin{minipage}{6.3pt}
\begin{picture}(6.2,28.5)
\put(1,13){\circle{4}}
\put(-2,0){2}
\end{picture}
\end{minipage}\;\;\;\longleftrightarrow \;\;\; v_{\sigma _1\sigma _2}
^{\alpha_1\alpha _2}(\tau _1,\tau _2):=\frac U2\delta (\tau _1-\tau _2)
\delta_{\sigma _1,-\sigma _2}\delta _{\alpha _1,-\alpha _2}\,.
\end{equation}
\end{mathletters}

With this graphical representation we can try to sum various
subclasses of diagrams either with or without renormalizations \cite{wor74}.
The renormalized quantities are standardly denoted with full circles and
bold lines, respectively. Note that in the present formulation of the LCE
the vertices contain the one-electron propagators (Green functions) while
the edges represent the electron-electron correlation.

To obtain a consistent LCE we must treat all the variational
functions and parameters that do not serve as generating sources in the
unperturbed, mean-field functional as external variables independent of the
generating fields. The mean-field stationarity equations, leading to an
approximate grand potential of the problem, are imposed only in the end of
the calculations after a chosen class of diagrams has been summed up and the
time-dependence from the generating fields removed. It is hence necessary to
have an expression for the grand potential to be able to expand consistently
around a MFT.

\section{Ring diagrams}

The LCE is of practical use only if we are able to pick up a class
of diagrams that can explicitly be summed and then numerically evaluated.
Due to the internal degrees of freedom (Matsubara frequencies) the quantum
version of the LCE is much more complicated than its classical counterpart.
We hence cannot expect to go that far in summing various classes of diagrams
as in the case of classical fluids or spin systems. In the quantum case we
are practically restricted to $\Phi $-derivable approximations \cite{note1}
where higher order vertices ($n> 2$) are neglected. We therefore resort to
LCE diagrams with {\em renormalized {\rm 1- and 2-vertices only, i.e.\ to
vertices that are at most second derivatives of the renormalized functional $%
g$. An approximation for the functional $g^{(N)}$ containing maximally $N$%
-vertices can be represented in the Baym form \cite{wor74}
\begin{equation}
\label{eq15}g^{(N)}=\exp \left\{ \sum_{n=1}^N\mbox{tr}\left[\widehat{C}_n
\widehat{\nabla}_x^n\right]\right\} g^{(0)}\{x\}\Big|_{x=0}+\Phi \left[
\widehat{\Gamma }\right] -\sum_{n=1}^N\mbox{tr}\left[\widehat{C}_n\widehat{%
\Gamma }_n\right]\,,
\end{equation}
where $\widehat{\Gamma }_n\,$ is a renormalized $n$-vertex, $\widehat{C}_n$
is a renormalized $n$-edge and $\widehat{\nabla}_x $ is a derivative in
appropriate generating sources $x$. Symbols with hats stand for matrices.
Here the trace tr is taken over all the internal degrees of freedom. The use
of generally renormalized quantities is crucial here, since only then the
functions $\widehat{\Gamma}_n$ and $\widehat{C}_n$ can be treated as
independent variational variables and the resultant approximation can lead
to a thermodynamically consistent and conserving approximation. The matrix
variables $\widehat{C}_n$ and $\widehat{\Gamma }_n$ are determined from
stationarity equations, i.e. $\delta g^{(N)}/\delta \widehat{\Gamma}
_n=\delta g^{(N)}/\delta \widehat{C}_n=0$.

If we restrict the theory to $N=2$ we are left with two
types of diagrams contributing to $\Phi $. The former is the effective-field
contribution involving only $\Gamma _1$. The functional $\Phi ^{(1)}$ then
takes the form
\begin{equation}
\label{eq16}\Phi ^{(1)}\left[ \Gamma \right] = \mbox{tr} \; \; \; \;
\begin{minipage}{23pt}
\begin{picture}(23.7,42.7)
\put(5.7,37){\circle*{11.5}}
\put(5.7,5.7){\circle*{11.5}}
\multiput(5.7,11.5)(0,8){3}{\line(0,1){4}}
\end{picture}
\end{minipage}
=-\frac{U}{2}\sum_\sigma \int \limits_0^\beta d\tau \, \frac{\delta g^{(1)}%
} {\delta \mu_\sigma (\tau)}\frac{\delta g^{(1)}}{\delta E_{-\sigma} (\tau)}%
{}.
\end{equation}
Since we expand around a mean field where $E_\sigma $ are variational
parameters, the derivative $\delta g^{(1)}/\delta E_{-\sigma }=0$ for
arbitrary $\tau \in [0,\beta ]$. The ``effective-field'' approximation
around a mean-field must hence vanish. This is a consequence of stationarity
of the mean-field approximation w.r.t.\ all variables.

A further step and the first nontrivial correction to a
MFT within the LCE are the ring diagrams consisting of single loops with
only 2-vertices $\widehat{\Gamma}_2$. It is easy to find a diagrammatic
representation for the ring diagrams contributing to the functional $\Phi$:

$$
\epsffile{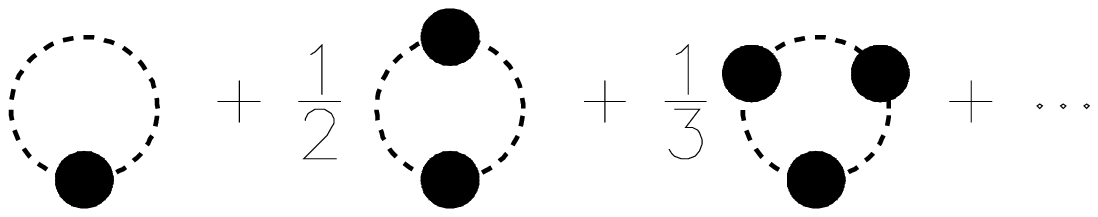}
$$

To find a mathematical expression for the sum of ring
diagrams we must first explicitly write down a matrix representation for
both the renormalized vertices (full circles) as well as for the bare edges
(dashed lines). They are $4\times 4$ matrices in the case of the
strong-coupling MFT and $2\times 2$ matrices in the weak-coupling MFT. In
the strong-coupling MFT we denote $\Gamma _{\sigma _1\sigma_2}^{\alpha
_1\alpha _2}=\delta ^2g^{(2)}/\delta x_{\sigma _1}^{\alpha _1}\delta
x_{\sigma _2}^{\alpha _2}$ and represent the 2-vertex as a matrix
\begin{equation}
\label{eq17}\widehat{\Gamma }_2=\left(
\begin{array}{cc}
\widehat{\Gamma}^{+ +}_{\sigma \sigma^{\prime}} & \widehat{\Gamma}^{+
-}_{\sigma \sigma^{\prime}} \\ \widehat{\Gamma}^{- +}_{\sigma
\sigma^{\prime}} & \widehat{\Gamma}^{- -}_{\sigma \sigma^{\prime}}
\end{array}
\right)
\end{equation}
where the single matrix elements are $2\times 2$ matrices. The bare edge $%
\widehat{v}$ can be represented as another $4\times 4$ matrix
\begin{equation}
\label{eq18}\widehat{v}=\frac U2\delta (\tau -\tau ^{\prime })\left(
\begin{array}{cc}
\widehat{0} & \widehat{\sigma}_1 \\ \widehat{\sigma}_1 & \widehat{0}
\end{array}
\right) \,,
\end{equation}
where $\widehat{\sigma}_1=\left(
\begin{array}{cc}
0 & 1 \\
1 & 0
\end{array}
\right)$ is the first Pauli matrix. Using this matrix representation we can
write down the sum of ring diagrams for the grand potential as
\begin{equation}
\label{eq19}\Phi ^{(2)}[\widehat{\Gamma} ]=-\frac 12\mbox{tr}\ln \left(
\widehat{1}- \widehat{v}\widehat{\Gamma}_2\right) \, .
\end{equation}
The prefactor $1/2$ compensates the doubling of contributions to the grand
potential due to the chosen matrix representation. Inserting (\ref{eq19}) in
(\ref{eq15}) we obtain
\begin{equation}
\label{eq20}g^{(2)}=\exp \left\{ \frac 12\mbox{tr}\left[\widehat{C}_2
\widehat{\nabla}_x^2 \right]\right\} g^{(0)}\left\{ x\right\} \Big|_{x=0}-%
\frac 12\mbox{tr}\ln \left( \widehat{1}-\widehat{v}\widehat{\Gamma}_2\right)
-\frac 12\mbox{tr} \left[\widehat{C}_2 \widehat{\Gamma}_2\right]
\end{equation}
where $\widehat{C}_2$ and $\widehat{\nabla}_x$ are also $4\times 4$
matrices. Approximation (\ref{eq20}) was used by Horwitz and Callen \cite
{hc61} and Bloch and Langer \cite{bl65} to improve the Weiss MFT for the
Ising model. Although solvable at the classical level, it is not viable at
the quantum level where the generating sources are time dependent. The
exponential in (\ref{eq20}) leads to a new functional integration being as
complicated as in the original problem. We hence have to approximate the
exponential. The simplest way to do that is to use the power expansion in $%
\widehat{C}_2$ and to terminate it appropriately. Since the most important
dynamical fluctuations come from second order perturbation theory, we resort
to an approximation reproducing the exact result asymptotically at least to $%
U^2$. A more complete treatment will be presented in Sec.\ VII.

To reproduce second order of perturbation theory it is
sufficient to take into account only the linear term in the operator $%
\widehat{C}_2$. Then (\ref{eq20}) reduces to
\begin{equation}
\label{eq21}g^{(2)}=g^{(0)}+\frac 12\mbox{tr}\left[\widehat{C}_2 \widehat{%
\nabla}_x^2\right]g^{(0)}\left\{ x\right\} \Big|_{x=0}-\frac 12 \mbox{tr}\ln
\left( \widehat{1}-\widehat{v}\widehat{\Gamma}_2\right) -\frac 12\mbox{tr}%
\left[\widehat{C}_2\widehat{\Gamma}_2\right]\,.
\end{equation}

Such an approximation is already explicitly solvable and
leads to an analytic expression for the grand potential $\Omega
^{(2)}=-\beta ^{-1}g^{(2)}$. We use the matrix representations for $\widehat{%
\Gamma}_2$ and $\widehat{v}$, (\ref{eq17}) and (\ref{eq18}), and evaluate
the second derivative of the unperturbed functional $g^{(0)}$. It
immediately follows from (\ref{eq8b}) that the derivatives with different
spin indices vanish and the matrix of the second derivative in (\ref{eq21})
substantially simplifies especially in the frequency representation.
Stationarity of the functional $g^{(2)}$ w.r.t.\ $\widehat{C}_2$ and $%
\widehat{\Gamma}_2$ yields explicit expressions for these matrix functions.
We exclude them from the potential $g^{(2)}$ and consider the paramagnetic
phase exclusively, so that the quantities loose their spin dependencies. We
then obtain for the physical grand potential
\begin{eqnarray}
\label{eq22}
&&\frac 1{2{\cal N}}\Omega _{MF}^{(2)}=-En^s-\beta
^{-1}\sum_{n=-\infty}^\infty e^{i\omega _n0^+}\left\{ \int\limits_{-\infty
}^\infty d\epsilon \rho (\epsilon )\ln \left[ i\omega _n+\mu-E
-\Sigma (i\omega_n)-\epsilon \right] +\ln G(i\omega _n)\right.\nonumber\\
&&\left. +(1-n^s)\left[ \ln \gamma (i\omega _n)
-\beta ^{-1}\ln \left( 1-n^s\right) \right] +n^s\left[
\ln \left[ \gamma (i\omega _n)-\frac U2\right] -\beta ^{-1}\ln
n^s\right] \right\}\nonumber\\
&&+\frac 1{4\beta}\sum_{\alpha =\pm 1}\sum_{m=-\infty}^\infty
e^{i\nu _m0_{}^{+}}\ln
\left[ 1-\alpha U\left\{ (1-n^s)\left\langle
G^{(0)}G^{(0)}\right\rangle (i\nu _m)+n^s\left\langle
G^{(U)}G^{(U)}\right\rangle (i\nu _m)\right\} \right] \,.
\end{eqnarray}
\thinspace Here $\omega _n=(2n+1)\pi \beta ^{-1}$, $\nu _m=2m\pi \beta ^{-1}$%
. We further used the abbreviations $\gamma(z)=G(z)^{-1}+\Sigma(z)$, $%
G^{(0)}(z)=\gamma (z) ^{-1}$, $G^{(U)}(z)=\left[ \gamma (z)-\frac U2\right]
^{-1}$ and
\begin{equation}
\label{eq23}\left\langle G^{(a)}G^{(a)}\right\rangle (z)=\beta
^{-1}\sum_{n=-\infty}^\infty G^{(a)}(i\omega _n)G^{(a)}(z+i\omega _n)\,
\end{equation}
with $a=0,U$. The mean-field parameters $E$ and $n^s$, and the complex
functions $\Sigma (i\omega _n)$ and $G(i\omega _n)$ at fixed $\mu $ are
treated in the approximate grand potential $\Omega ^{(2)}$ variationally in
the same way as in the unperturbed theory with $\Omega ^{(0)}$. The new
grand potential $\Omega ^{(2)}$ defines then a thermodynamically consistent
theory.

An analogous expression can be derived for the expansion
around the weak-coupling MFT. Explicitly we obtain
\begin{eqnarray}
\label{eq24}
\frac 1{2{\cal N}}\Omega _{HF}^{(2)}&=&-\frac U2n^2-\beta ^{-1}\sum_{n=-\infty
}^\infty e^{i\omega _n0^+}\left\{ \int\limits_{-\infty }^\infty d\epsilon \rho
(\epsilon )\ln \left[ i\omega _n+\mu-Un-\Sigma (i\omega _n)-\epsilon \right]
+\ln G(i\omega _n)\right.\nonumber\\
&&\left. +\ln \gamma (i\omega _n) \right\}
+\frac 1{4\beta }\sum_{\alpha =\pm 1}\sum_{m=-\infty}^\infty
e^{i\nu _m0_{}^{+}}\ln \left[
1-\alpha U\left\langle G^{(0)}G^{(0)}\right\rangle (i\nu _m)\right]
\end{eqnarray}
where the function $G^{(0)}$ has the same meaning as in (\ref{eq22}). Grand
potentials (\ref{eq22}) and (\ref{eq24}) hold only for the paramagnetic
phase. Grand potentials allowing for a broken symmetry can be derived
without conceptual changes.

Having explicit expressions for the contributions due to
the ring diagrams we can compare the resulting theory with other similar
approximate schemes, especially those studied extensively by Menge and
M\"uller-Hartmann \cite{mm91}, iterated perturbation theory (IPT) introduced
recently \cite{gk92,zrk93,gkr93} in connection with the metal-insulator
transition in the $d=\infty $ Hubbard model, and with the extended
Edwards-Hertz approximation \cite{eh91,wc94}. This will be done in Secs.\ VI
and VII. Before we do this we should realize that the presented sum of
diagrams within the LCE scheme does not renormalize the one-electron Green
function but rather the interaction strength. The ring-diagram approximation
(\ref{eq22}), (\ref{eq24}) represents a sum of a geometric series of
2-particle bubbles renormalizing the interaction, i.e.\ diagrams
contributing to the polarization of the vacuum. Because of the locality of
the electron-electron interaction in the Hubbard model only even powers of $U
$ contribute. Grand potentials (\ref{eq22}), (\ref{eq24}) correspond to an
``effective-interaction'' (bubble-chain) approximation studied at the level
of self-energy in \cite{mm91}. However, the derived sum of ring diagrams
does not represent a fully self-consistent approximation, since the local
electron propagators in the expansion are unrenormalized, mean-field ones, $%
G^{(0)},G^{(U)}$. Although these propagators contain the self-energy $\Sigma
$ and are not free, the dependence of $G^{(a)}$ on $\Sigma $ is caused by
the reduction of a $d=\infty $ theory to a $d=0$ (single-site) problem. Such
a partial self-consistency is of topological rather than of dynamical
origin. Moreover, neither (\ref{eq22}) nor (\ref{eq24}) has the Baym form
\cite{ba62} where the $\Phi$-functional depends only on the full propagator $%
G$. We hence have no guarantee that they are conserving approximations.

\section{One-particle properties}

Grand potential (\ref{eq22}) leads to a numerically
solvable theory. In the present form it is only suitable for the
paramagnetic phase and we use it to evaluate explicitly the one-electron
properties. In order to simplify the numerical procedure we restrict our
considerations to zero temperature.

The defining equations for the variational variables are
derived from stationarity of $\Omega ^{(2)}$. Vanishing of the derivatives
w.r.t.\ the static variables $n^s$ and $E$ leads to
\begin{mathletters}
\label{eq25}
\begin{equation}
\label{eq25a}
E={\cal E}+\frac U{4\beta }\sum_{m=-\infty}^\infty e^{i\nu _m0^{+}}
\sum_{\alpha=\pm 1}
\alpha C_\alpha (i\nu_m)\left[\left\langle G^{(0)}G^{(0)}\right\rangle
(i\nu _m)-\left\langle
G^{(U)}G^{(U)}\right\rangle (i\nu _m)\right]\; ,
\end{equation}
\begin{equation}
\label{eq25b}
n^s=\beta ^{-1}\sum_{n=-\infty}^\infty e^{i\omega _n0^+}\int\limits_{-
\infty }^\infty d\epsilon \rho (\epsilon )\left[ i\omega _n+\mu-E-\Sigma
(i\omega _n)-\epsilon \right] ^{-1}\, ,
\end{equation}
where
\begin{equation}
\label{eq25c}
{\cal E}=-\beta^{-1}\sum_{n=-\infty}^\infty e^{i\omega_n0^+}\ln
\left[1-\frac U2G^{(0)}(i\omega_n)\right]
\end{equation}
and
\begin{equation}
\label{eq26c}
C_\alpha (z)=\left[ 1-\alpha U\left\{ (1-n^s%
)\left\langle G^{(0)}G^{(0)}\right\rangle (z)+n^s\left\langle
G^{(U)}G^{(U)}\right\rangle (z)\right\} \right] ^{-1}\,.
\end{equation}
\end{mathletters}
The term in (\ref{eq25a}) including a sum over bosonic frequencies is a
correction to the unperturbed static MFT due to the ring diagrams.
Variations of $\Omega ^{(2)}$ w.r.t.\ $\Sigma (i\omega _n)$ and $G(i\omega
_n)$ lead to corrected mean-field equations for these functions:
\begin{mathletters}
\label{eq26}
\begin{equation}
\label{eq26a}
G(i\omega _n)=\int\limits_{-\infty }^\infty d\epsilon \rho
(\epsilon )\left[ i\omega _n+\mu-E-\Sigma (i\omega _n)-\epsilon \right] ^{-1}
\end{equation}
and%
\begin{eqnarray}
\label{eq26b}
&G(i\omega _n)&=(1-n^s)G^{(0)}(i\omega _n)+n^s%
G^{(U)}(i\omega _n)\nonumber\\
&&-\frac U{4\beta }(1-n^s)G^{(0)}(i\omega _n)^2\sum_{\alpha=\pm 1} \alpha
\sum_{m=-\infty}^\infty e^{i\nu _m0^{+}}
 G^{(0)}(i\omega _n+i\nu _m)\left[ C_\alpha (i\nu_m)
+C_\alpha (-i\nu _m)\right]\nonumber\\
&&-\frac U{4\beta }n^sG^{(U)}(i\omega _n)^2\sum_{\alpha=\pm 1}
\alpha \sum_{m=-\infty}^\infty e^{i\nu _m0^{+}}
G^{(U)}(i\omega _n+i\nu _m)\left[ C_\alpha(i\nu _m)+C_\alpha (-i\nu _m)
\right]\, .
\end{eqnarray}
\end{mathletters}

To facilitate the numerical evaluation of the derived
equations we convert the sums over Matsubara frequencies to integrals over
real frequencies. However, this is possible only if the functions under
consideration are analytic in the upper and lower complex half-planes. Apart
from the one-particle Green function, the analytic structure of which is
clear, there are two other complex functions $\left\langle
G^{(a)}G^{(a)}\right\rangle (z)$ and $C_\alpha (z)$. The former function is
analytic with a cut along the real axis as can be seen from the
representation
\begin{equation}
\label{eq27}X^{(a)}(z):=\left\langle G^{(a)}G^{(a)}\right\rangle (z)= -\frac
1\pi\int\limits_{-\infty }^0dx\left[ G^{(a)}(x+z)+G^{(a)}(x-z)\right]
\mbox{Im}G^{(a)}(x+i0^{+})\,.
\end{equation}
On the other hand the function $C_\alpha(z)$ can have a pole in the complex
plane if
\begin{equation}
\label{eq28}1=\alpha U\left\{ (1-n^s)X^{0}(z)+n^s X^{(U)} (z)\right\} \,.
\end{equation}
Since $X^{(a)}(z)=X^{(a)}(-z)$, the r.h.s.\ of (\ref{eq28}) is real for $z=iy
$, i.e.\ along the imaginary axis. There may be a critical value $U_c$ at
which the pole appears for the first time at $z=0$ and where the
perturbation expansion breaks down. We must hence distinguish two regimes of
the interaction strength. If $U<U_c$ (weak coupling) we have no polar
contributions to the spectral representations of the Matsubara sums. At
strong coupling, $U>U_c$, the pole in $C_\alpha(z)$ influences the integrals
over real frequencies. The existence of a pole in a two-particle Green
function indicates an instability of perturbation theory and is connected
with a metal-insulator transition. We postpone a detailed investigation of
the actual role of such a pole to a separate publication and stick in this
paper exclusively to the weak-coupling regime $U<U_c$. In this case it is
easy to show that the sum over bosonic frequencies in (\ref{eq25a}) vanishes
due to analyticity of the integrand in the upper and lower complex
half-planes. Therefore the mean-field equations for $n^s$ and $E$ are not
changed by the contributions from the ring diagrams. Eq. (\ref{eq26b}) can
be rewritten at weak coupling for an arbitrary complex energy $z$ as
\begin{mathletters}
\begin{equation}
\label{eq29}
G(z)=(1-n^s)G^{(0)}(z)\left\{ 1+\frac U2G^{(0)}(z)I^{(0)}(z)\right\}
+n^sG^{(U)}(z)\left\{1+\frac U2G^{(U)}(z)I^{(U)}(z)\right\} \,,
\end{equation}
where we used
\begin{equation}
\label{eq31b}
I^{(a)}(z)=\sum_{\alpha=\pm 1} \frac \alpha\pi \int\limits_{-\infty }^0dx
\left[ C_\alpha (z-x)\mbox{Im}G^{(a)}(x+i0^{+})-G^{(a)}(z+x)\mbox{Im}C_\alpha
(-x+i0^{+})\right] \,.
\end{equation}
\end{mathletters}

The function $C_\alpha (z)$ is defined through the
functions $X^{(a)}$ as
\begin{equation}
\label{eq30}C_\alpha (z)=\left\{ 1-\alpha U\left[
(1-n^s)X^{(0)}(z)+n^sX^{(U)}(z)\right] \right\} ^{-1}\,.
\end{equation}
Equation (\ref{eq29}) can be turned into an equation for the self-energy $%
\Sigma (z)$ if we use the definition of $G^{(a)}$ given below eq.\ (\ref
{eq22}). We then obtain
\begin{eqnarray}
\label{eq31}
\Sigma(z)&=&\frac U2\frac{n^s}{1+G(z)\left[ \Sigma(z)
-U/2\right] }\left\{ 1+\frac{1+G(z)\Sigma(z)}{1+G(z)\left[ \Sigma(z)-U/2
\right]}I^{(U)}(z)\right\}  \nonumber\\[8pt]
&&+\frac U2\frac{1-n^s}{1+G(z)\Sigma(z)}I^{(0)}(z)\,.
\end{eqnarray}
Eq.\ (\ref{eq31}), completed with the mean-field equations
\begin{mathletters}
\label{eq32}
\begin{equation}
\label{eq32a}
E=\frac 1\pi \int\limits_{-\infty }^0dx\mbox{Im}\ln \left[ 1-\frac
U2G^{(0)}(x+i0^{+})\right]
\end{equation}
and
\begin{equation}
\label{eq32b}
n^s=-\frac 1\pi \int\limits_{-\infty }^0dx \mbox{Im}G(x+i0^+) \,,
\end{equation}
\end{mathletters}
is to be solved numerically for $z=\omega +i0^{+}$. From (\ref{eq32b}) we
see that $n^s=n$ as is the case in the static MFT from \cite{vjdv92}. Due to
the electron-hole symmetry it is sufficient to solve the equations for $%
\omega \leq 0$. To obtain the values for the self-energy and the
one-electron Green function on the positive frequency axis we use the
electron-hole transformation ($n\to 1-n, \omega\to -\omega$) leading to:
\begin{mathletters}
\label{eq33}
\begin{equation}
\label{eq33a}
G(-\omega+i0^{+})=-G(\omega+i0^{+})^{*}
\end{equation}
and
\begin{equation}
\label{eq33b}
\Sigma (-\omega+i0^{+})=\frac U2-\Sigma (\omega+i0^{+})^{*}\,.
\end{equation}
\end{mathletters}
The chemical potential transforms in this expansion as $\mu\to U/2-\mu$ and
the parameter $E\to -E$. It is to be proved that (\ref{eq33}) really holds
for a solution of (\ref{eq31}), i.e.\ the approximate theory correctly
exhibits the electron-hole symmetry. The proof is presented in the Appendix.

The same steps can be made to derive equations for the
self-energy of the ring diagrams summed around the Hartree-Fock grand
potential (\ref{eq24}). The resulting self-energy then is
\begin{equation}
\label{eq34}\Sigma _{HF}(z)=\frac{U/2}{1+G(z)\Sigma (z) }I^{(0)}(z)\,.
\end{equation}
We see that the self-energy from the strong-coupling MFT formally reduces to
$\Sigma_{HF}$ if we put $n^s=0$ and $E=Un$.

The electron-hole symmetry in the weak-coupling case takes
the form
\begin{mathletters}
\label{eq35}
\begin{equation}
\label{eq35a}
G(-\omega+i0^{+})=-G(\omega+i0^{+})^{*}\, ,
\end{equation}
\begin{equation}
\label{eq35b}
\Sigma _{HF}(-\omega+i0^{+})=-\Sigma _{HF}(\omega+i0^{+})^{*}\,.
\end{equation}
\end{mathletters}

\section{Numerical results and comparison of
various approximate schemes}

We now apply the derived equations in the weak-coupling
limit to assess quantum fluctuations due to dynamical corrections to the
static MFT. Only for the sake of simplicity we perform the numerical
calculations on a Bethe lattice in $d=\infty $, i.e.\ we use for the
diagonal element of the one-electron Green function the representation
\begin{equation}
\label{eq36}G^{(0)}(z)=\left[ z-\frac{D^2}4G(z)\right] ^{-1}
\end{equation}
where $D=2$ equals half the bandwidth. We also restrict the calculation to
the half-filled band in the paramagnetic phase, i.e.\ $n_\sigma =n_\sigma
^s=1/2$ and $\mu =U/4$ $\,(\mu -E=0)$. Furthermore we remain during all the
calculations at $T=0$.

Since our aim is to investigate the resulting equations
only in the weak-coupling limit, we separately analyze the $U^2$ corrections
to the static MFT before we take into account all the ring diagrams. In
course of calculations we compare the results obtained from the LCE treated
here with other approximations of similar origin.

\subsection{$U^2$-corrections to the mean-field solutions}

Let us start with the simpler expansion around the
Hartree-Fock MFT. Using (\ref{eq34}) we obtain for the $U^2$-contributions
to the self-energy
\begin{eqnarray}
\label{eq37}
\Sigma (z)&=&\frac{U^2}{1+G(z)\Sigma (z)}I_0^{(0)}(z)\equiv \frac{U^2}{1+G(z)
\Sigma (z)}\int\limits_{-\infty }^0\frac{dx }\pi \left[
X^{(0)}(z-x )\mbox{Im}G^{(0)}(x +i0^{+})\right. \nonumber\\
&&\left. +G^{(0)}(z+x)\mbox{Im}X^{(0)}(x +i0^{+})\right] \,.
\end{eqnarray}
The resulting self-energy consists of two different contributions, the
algebraic prefactor and the integral $I_0^{(0)}$. The former is a
consequence of the incomplete self-consistency of the defining equation (\ref
{eq37}). The Green function $G^{(0)}$ from the r.h.s.\ of (\ref{eq37}) is
not identical with the full one-electron propagator $G$. Hence the variation
of the grand potential (\ref{eq22}) w.r.t.\ $G$, leading to the equation for
the self-energy, generates algebraic factors due to the variation of the
propagator $G^{(0)}$. The latter contribution to (\ref{eq37}) is exactly the
result of the iterated perturbation theory (IPT) and corresponds to a
partially self-consistent, second-order Feynman diagram for the self-energy.
It is important that both the contributions are complex and lead to a rather
intricate analytic structure of the self-energy $\Sigma (z)$. It is
impossible to prove the desired Herglotz properties, i.e.\ $\Sigma
(z)=\Sigma (z^{*})^{*}$ and $\mbox{Im}\Sigma (z) \propto -\mbox{Im}z$ for
all complex energies $z$. Only at the Fermi level, $\omega =0$, we obtain
from the electron-hole symmetry that $\Sigma (0)=0$.

Approximation (\ref{eq37}) does not seem to cause troubles
at weak coupling ($U\leq 1.9$), where only a slight modification of the
non-interacting density of states (DOS) can be observed (Fig.\ 1). However,
interactions stronger than $U_c\approx 1.9$ lead to a breakdown of the
approximate equations, since a singularity appears in the prefactor $\left[
1+G(z)\Sigma (z)\right] ^{-1}$. Namely we observe that the derivative $%
\partial G^{(0)}(\omega)/\partial\omega$ diverges at a critical frequency $%
\omega_c$, which leads to a jump in the propagator $G^{(0)}(\omega)$. The
density of states then develops a cusp at the critical frequencies $\pm
\omega_c$ (Fig.\ 2). In such a situation it is impossible to reach
convergence and it is unclear if the equation (\ref{eq37}) does have a
solution at all.

The algebraic prefactor is a consequence of the variation
of the grand potential and has no diagrammatic background. It causes the
breakdown of the iteration scheme. We may ask if we can do better by
neglecting this prefactor. Such a choice corresponds to the diagrammatic
expansion directly applied to the self-energy instead of to the grand
potential. We obtain
\begin{equation}
\label{eq38}\Sigma (z)=U^2\int\limits_{-\infty }^0\frac{dx}\pi \left[
X^{(0)}(z-x)\mbox{Im}G^{(0)}(x+i0^{+})+G^{(0)}(z+x)\mbox{Im}%
X^{(0)}(x+i0^{+})\,\right] ,
\end{equation}
which is the IPT self-energy. Analyzing (\ref{eq38}) we find that this
equation can safely be iterated up to $U\approx 4.7$ (Fig.\ 3) where the
results of Georges and Krauth \cite{gkr93} are reproduced. The DOS shows a
three-peak structure where the central band narrows with increasing $U$.
However, it seems impossible to reach the alleged metal-insulator transition
at which the width of the middle band has to vanish. Before we reach this
point, the iterations break down for similar reasons as in the solution of (%
\ref{eq37}). With increasing interaction strength a discontinuity in the
real part of the Green function $G^{(0)}(\omega )$ seems to develop together
with a cusp in its imaginary part (Figs.\ 4a, b). The extremal points for
both the functions approach each other (Fig.\ 5) and seem to merge at $%
\omega _c<0$, i.e.\ away from the Fermi energy. Earlier a singularity in $%
G^{(0)}$ was reported on the real axis only at the Fermi energy \cite{zrk93}%
. According to our results the Green function $G^{(0)}$ displays a
discontinuity at $\omega _c$ and it is again impossible to reach a solution
via iterations. If we attempt to solve (\ref{eq38}) for complex energies,
the breakdown of the iteration scheme is shifted to stronger interactions
and the critical frequency is pushed towards the Fermi energy. Although the
singularity smooths in comparison with the real frequencies, we did not
succeed to remove it completely from $G^{(0)}$ and to reach a convergent
solution at strong coupling. If there were a singularity at $G^{(0)}$ even
at complex frequencies, the Green function could not fulfil the desired
Herglotz properties. However, if the Herglotz properties of $G^{(0)}(z)$
were violated, it would be impossible to continue analytically the
quantities from the imaginary frequency axis to the real energies in a
unique way. The approximation then would become unreliable with possible
spurious poles (zeros) in the complex plane. It is hence of great importance
now to prove or disprove the Herglotz properties of the IPT self-energy if
we want to draw conclusions from this approximation especially at strong
coupling.

{}From the above results it may seem that the IPT for the
self-energy be superior to the LCE for the grand potential because of the
unphysical complex prefactor in (\ref{eq37}). This prefactor appears due to
the fact that the one-electron propagator $G^{(0)}$, used in the ring
diagrams, deviates from the full propagator $G\,$ and thus the expansion is
not fully self-consistent. It is a consequence of the non-self-consistent
character of the expansion that not all physical quantities have a
diagrammatic expansion. Diagrammatically the derivative is not connected
with a cut of the one-electron propagator only, but a variation of $G^{(0)}$
w.r.t. $G$ appears. This variation leads to the lost of the diagrammatic
control over physical quantities and causes violation of Ward identities.
Undesired prefactors are inevitable companions of non- or only partly
self-consistent theories. Although nonexistent at the level of self-energy
within the IPT, the unphysical prefactors do appear in higher order Green
functions and correlation functions (susceptibilities). Therefore we cannot
rely on IPT susceptibilities \cite{gkr93}. Since the existence of
``spurious'' complex prefactors is a ubiquitous feature of partly
self-consistent theories, we may conclude that only fully self-consistent
theories in the spirit of Baym offer a physically consistent global picture.

We can also calculate the $U^2$ corrections to the
strong-coupling MFT. There we meet the same problems with prefactors and a
breakdown of the iteration scheme as in the case of the expansion around the
Hartree-Fock MFT.

\subsection{Corrections to the mean-field
solutions due to the ring diagrams}

One may still hope to improve the weak-coupling
approximation and to avoid the breakdown of the iteration scheme by taking
into account the ring diagrams. However, the theory with ring diagrams
suffers from the same deficiency as the $U^2$ terms, namely the equations
contain unphysical complex prefactors, and an analogous breakdown of the
iteration procedure is to be expected. It turns out that for the ring
diagrams a singularity in the prefactor $\left[ 1+G(z)\Sigma (z)\right] ^{-1}
$ shows up even for smaller values of the interaction, namely $U\approx 1.66$%
, where the approximation scheme breaks down. Above this value the
iterations do not converge any longer. It indicates that the additive
diagrams push down the ``transition'' from the weak to the strong coupling
regimes. To confirm this we suppressed the prefactors and thus gained an
extension of the self-energy from the IPT (\ref{eq38}) due to the ring
diagrams. The result is plotted in Fig.\ 6. We can see that the self-energy
has a narrow central peak with a tendency to build up two satellite bands.
However, before the satellites can be fully formed a pole in the
two-particle function $C_\alpha(\omega)$ (\ref{eq30}) at $\omega=0$ is
reached at $U\approx 1.9$ and the theory breaks down. There is no way to go
around the pole in the two-particle function $C_\alpha(z)$ in this not fully
self-consistent approximation. The pole causes the Herglotz properties to be
manifestly broken. There is no convergent solution for $U>1.9$. The only way
to restore analytic properties of the local, one-electron Green function is
to make the theory fully self-consistent.

\section{Self-consistencies and renormalizations
in the linked cluster expansion}

We saw in the preceding section that non-self-consistent
diagrammatic approximations lead to differences in physical quantities
derived either diagrammatically or with the use of derivatives (variations)
of the grand potential w.r.t.\ external sources. This is a general feature
of not fully self-consistent approximations where some Ward identities are
violated. This conclusion applies also to the approximations investigated in
Secs.\ IV-VI. Although the corresponding grand potential has a form where
all the unknown functions have to be determined from stationarity
conditions, the internal electron propagators in the diagrams are $%
G^{(0)}=(G^{-1}+\Sigma )^{-1}$ (or $G^{(U)}=(G^{-1}+\Sigma -U/2)^{-1}$)
instead of $G$ as one expects in the Baym approach. These are dynamically
free propagators, since the dependence on the self-energy does not stem from
the dynamics but rather from a special topology of the $d=\infty $ lattice
mapped onto a single site (impurity) embedded in a medium. E.g.\ if the
diagrams were treated in $d=3$, the function $G^{(0)}(z)$ had to be
replaced by the bare (unperturbed) ${\bf k}$-dependent propagator $G^{(0)}(%
{\bf k},z)$. There would be no self-consistency at all in finite dimensional
lattices. The origin of the partial self-consistency in the investigated
approximate schemes lies entirely in the reduction of the topology of the $%
d=\infty $ model to a ``self-consistent'' single-site  problem
\cite{vjdv92a}. The topological self-consistency in the local
propagator $G^{(0)}(z)$ coherently copies the dynamical processes from the
impurity to the medium which, on the other hand, provides the electrons
for the impurity.

In this light it is clear that e.g.\ the self-energy,
obtained by varying the grand potential from the LCE, does not equal the
self-energy obtained from a sum of the non-self-consistent self-energy
diagrams. To get rid of such ambiguity problems we must abandon
non-self-consistent theories and introduce true dynamical renormalizations
into the LCE. We can try to introduce the self-consistency naively via a
replacement of the unperturbed electron propagator $G^{(0)}$ with the full
propagator $G$. However, such an introduction of the self-consistency into
the theory is dangerous, since it need not be diagrammatically controlled.
This naive way of turning the theory self-consistent may lead to multiple
counting of diagrams and to undesired unphysical and spurious structures.
The only consistent way to introduce self-consistencies into diagrammatic
expansions is to classify diagrams into particular classes which may be
summed by iterations, i.e.\ by self-similar insertions.

Possible renormalization schemes within the LCE were
extensively discussed by Wortis \cite{wor74}. However, in the quantum case
with internal time (frequency) degrees of freedom, we are much more
restricted to produce a tractable approximation. As we already mentioned,
even a quantum analogue of the Bloch-Langer approximation from the classical
Heisenberg model \cite{bl65} is not viable in the quantum case. Due to the
time dependence of the vertices the ordering of functional derivatives,
i.e.\ the time ordering of the external legs at vertices, is important.
Interchange of derivatives leads to ``crossing'' of interaction lines and to
different quantitative results for diagrams with vertices of higher order
than two. If we resort to a fixed time ordering of the vertex legs, i.e.\ to
a kind of ``noncrossing approximation'' we obtain a self-consistent version
of the ring-diagram approximation.

To this end we must take  into account noncrossing
diagrams resulting from the expansion of the exponential $\exp \left\{ \frac
12\mbox{tr}\left[\widehat{C}_2 \widehat{\nabla}_x^2 \right]\right\} $ in (%
\ref{eq20}). If we analyze the contributions from products of the operator $%
\widehat{C}_2\widehat{\nabla}_x^2$ we can classify them into crossing and
noncrossing diagrams, according to whether the two-point function $C_2$ is
or is not crossed by an interaction line (Fig.\ 8), respectively. The
topology of higher order diagrams becomes more and more intricate due to the
crossed diagrams (Fig.\ 8d). If we neglect them we can exactly sum up the
contributions from the noncrossing diagrams via self-consistency. Insertions
as in Fig.\ 8c can be summed by replacing $G^{(0)}$ with $\overline{G}$,
where the self-energy due to the $C$-contributions (hatched bubbles) equals
to a convolution $\overline{G}*C$. The grand potential due to the
self-consistent ring diagrams around the Hartree-Fock MFT can then be
written as a generating Baym functional%
\begin{eqnarray}
\label{eq39}
&&\frac 1{{\cal N}}\Omega _{HF}^{(ring)}=-U n_{\uparrow }
n_{\downarrow }
-\beta ^{-1}\sum_\sigma \sum_{n=-\infty}^\infty e^{i\omega _n0^{+}}\left\{
\int\limits_{-\infty }^\infty d\epsilon \rho (\epsilon )
\ln \left[ i\omega _n+\mu
-Un_{-\sigma }-\Sigma _\sigma (i\omega _n)\right.\right.\nonumber\\
&&\left.\left.-\Delta \Sigma _\sigma (i\omega_n)-\epsilon \right]
+\Delta \Sigma_\sigma (i\omega _n)
\overline{G}_\sigma (i\omega _n)
+\ln \left[ 1+G_\sigma (i\omega _n)\left( \Sigma _\sigma (i\omega _n)-\langle%
\overline{G}_\sigma C_\sigma \rangle (i\omega _n)\right) \right]\right\}
\nonumber\\
&&+\beta ^{-1}\sum_{m=-\infty}^\infty e^{i\nu _m0^{+}}\left\{ \sum_\sigma
\Gamma_\sigma (i\nu _m)C_\sigma (i\nu _m)+\frac 12\ln \left[ 1-U^2\Gamma
_{\uparrow }(i\nu _m)\Gamma _{\downarrow }(i\nu _m)\right] \right\}
\end{eqnarray}
where we used the same notation for the convolution $\langle\overline{G}%
_\sigma C_\sigma \rangle(i\omega _n)$ as in (\ref{eq23}), but with the sum
over bosonic Matsubara frequencies. Here $\Delta \Sigma _\sigma $ is a
correction due to the $C_\sigma$-induced contributions from the noncrossing
diagrams and $\overline{G}_\sigma $ is its Legendre conjugate variable. If
they vanish, (\ref{eq39}) reduces to its non-self-consistent version (\ref
{eq24}). If we exclude the variables $\Gamma_\sigma$, $C_\sigma$ and $%
\overline{G}_\sigma$, $\Delta \Sigma_\sigma $ we can reduce (\ref{eq39}) at
the saddle point to
\begin{eqnarray}
\label{eq40}
&&\frac 1{{\cal N}}\Omega _{HF}^{(ring)}=-U n_{\uparrow }n_{\downarrow}
-\beta ^{-1}\sum_\sigma \sum_{n=-\infty}^\infty e^{i\omega _n0^{+}}\left\{
\int\limits_{-\infty }^\infty d\epsilon \rho (\epsilon )\ln \left[
i\omega _n+\mu -Un_{-\sigma }-
\Sigma_\sigma(i\omega _n)-\epsilon \right] \right.\nonumber\\
&& \left. + G_\sigma (i\omega _n)\Sigma_\sigma(i\omega _n)\right\}
+\frac 1{2\beta }\sum_{m=-\infty}^\infty e^{i\nu _m0^{+}}\ln \left[
1-U^2\langle G_{\uparrow }G_{\uparrow }\rangle(i\nu _m)
\langle G_{\downarrow }G_{\downarrow}\rangle (i\nu _m)\right] \,.
\end{eqnarray}
Grand potential $\Omega _{HF}^{(ring)}$ is the fully self-consistent
``bubble-chain'' approximation investigated at the level of self-energy by
Menge and M\"uller-Hartmann \cite{mm91}.

Analogously to (\ref{eq39}) we derive a self-consistent
version of the ring-diagram grand potential for the LCE around the
strong-coupling MFT:%
\begin{eqnarray}
\label{eq41}
\frac{1}{{\cal N}}\Omega _{MF}^{(ring)}&=&
-\beta ^{-1}\sum_\sigma
\sum_{n=-\infty}^\infty e^{i\omega _n0^{+}}\left\{\int\limits_{-\infty }^
\infty d\epsilon \rho (\epsilon )\ln \left[ i\omega_n+\mu -E_\sigma -
\Sigma _\sigma (i\omega _n)-\Delta \Sigma _\sigma (i\omega_n)-\epsilon
\right] \right.
\nonumber \\
&&\left.+\left( 1-n^s_{-\sigma }\right) \ln \left[ 1+G_\sigma
(i\omega _n)\left( \Sigma _\sigma (i\omega _n)-\langle\overline{G}_\sigma
C_\sigma\rangle(i\omega _n)\right) \right]\right.
+ \Delta\Sigma _\sigma (i\omega _n)\overline{G}_\sigma (i\omega _n)
\nonumber\\
&&\left. +n^s_{-\sigma }\ln \left[ 1+G_\sigma (i\omega _n)\left(
\Sigma _\sigma (i\omega _n)-\frac U2-\langle\overline{G}_\sigma C_\sigma
\rangle(i\omega_n)\right) \right] \right\}-\sum_\sigma E_\sigma n^s_\sigma
\nonumber\\
&&+\beta ^{-1}\sum_{m=-\infty}^\infty e^{i\nu _m0^{+}}\left\{ \sum_\sigma
\Gamma_\sigma (i\nu _m)C_\sigma (i\nu _m)+\frac 12\ln \left[ 1-U^2\Gamma
_{\uparrow }(i\nu _m)\Gamma _{\downarrow }(i\nu _m)\right] \right\} \,.
\end{eqnarray}
Excluding the variables $\Gamma _\sigma $ and $C_\sigma $ at the saddle
point we obtain defining equations for the Green functions $G_\sigma $ and $%
\overline{G}_\sigma $%
\begin{mathletters}
\label{eq42}
\begin{equation}
\label{eq42a}
\overline{G}_\sigma (i\omega _n)=G_\sigma (i\omega
_n)=\int\limits_{-\infty }^\infty d\epsilon \rho (\epsilon )\frac 1{i\omega
_n+\mu -E_\sigma -\Sigma _\sigma (i\omega _n)-\Delta \Sigma _\sigma (i\omega
_n)-\epsilon }
\end{equation}
and for the self-energies $\Sigma _\sigma (i\omega _n)$ and $\Delta \Sigma
_\sigma (i\omega _n)$
\begin{equation}
\label{eq42b}
\Delta \Sigma _\sigma (i\omega _n)=-\frac{U^2}{2\beta }%
\sum_{m=-\infty}^\infty G_\sigma (i\omega _n+i\nu _m)\frac{\langle G_{-
\sigma }G_{-\sigma }\rangle(i\nu _m)}{1-U^2\langle G_{\uparrow }G_{\uparrow
}\rangle(i\nu _m)\langle G_{\downarrow }G_{\downarrow}\rangle (i\nu _m)}\,,
\end{equation}
\begin{equation}
\label{eq42c}
\Sigma _\sigma (i\omega _n)-\Delta \Sigma _\sigma (i\omega _n)=%
\frac U2\frac{n^s_{-\sigma }}{1+G_\sigma (i\omega _n)\left( \Sigma
_\sigma (i\omega _n)-\Delta \Sigma _\sigma (i\omega _n)-U/2\right) }\,.
\end{equation}
\end{mathletters}
Grand potential $\Omega^{(ring)}_{MF}$ has an attractive feature in that it
contains the split-band limit of the Hubbard-III solution and is exact at
weak coupling up to $U^2$. In contrast to previously studied theories with
analogous properties \cite{eh91,wc94}, grand potential (\ref{eq41}) is
derived in a fully controlled manner. There are no spurious prefactors in
the defining equations for the self-energies calculated from (\ref{eq39})
and (\ref{eq41}). There is also no artificial breakdown of the iteration
scheme in the fully self-consistent theory. All the Ward identities are now
fulfilled and there is no difference in deriving various physical quantities
either from varying the grand potential or using the corresponding
diagrammatic expansion for them, since the theory is thermodynamically
consistent and conserving in the Baym sense.

We can now analytically continue the functions $%
\Sigma_\sigma$, $\Delta \Sigma_\sigma$ and $G_\sigma$ to the whole complex
plane and use spectral representations with real frequencies. The numerical
solution is then obtained straightforwardly. The quantitative results in the
paramagnetic phase (Fig.\ 8) obtained from (\ref{eq40}) at weak coupling ($%
U\leq 3.5$) agree with the results of the ``bubble-chain'' approximation
from \cite{mm91}. For stronger coupling a pole in the two-particle function $%
C_\sigma(\omega)$ interferes and the iteration scheme must be modified to
preserve the Herglotz properties of the one-electron propagator \cite{note2}.

Self-consistent mean-field theory with dynamical
fluctuations (\ref{eq42}) exhibits similar behavior at weak coupling ($U\leq
1.5$) (Fig.\ 9). The DOS at the Fermi energy begins to decrease, since the
theory is not a Fermi liquid (the term proportional to $U^3$ violates the
Luttinger theorem). Two satellite peaks start to develop and split off at
about $U_c\approx 5.5$ ($U_c=4$ for the static MFT) similarly to \cite{wc94}%
. Because of frequency convolutions there are no sharp (algebraic) band
edges and it is numerically hard to determine when exactly the bands split
off. We also had troubles to reach convergence for the frequencies lying
within the expected gap. To reach a reliable solution in the insulating
phase, the iteration scheme must be refined. We, however, did not meet any
singularities indicating an instability of this dynamical mean-field theory.
The two-particle function $C_\sigma(\omega)$ did not reach the pole (\ref
{eq28}) because of the metal-insulator transition and we did not observe any
indication for a divergency in the perturbation theory.

The self-energy from (\ref{eq42}) contains two parts, $%
\Delta \Sigma $ being the direct contribution of the ring diagrams and $%
\Sigma -\Delta \Sigma $ which is the mean-field self-energy with the
renormalized one-electron propagator. To assess these two different
self-energies we plotted their imaginary parts separately in Figs.\ 10a, b
and compared them with the corresponding self-energies from the Hartree-Fock
solution (\ref{eq40}) and the static MFT \cite{vjjmdv93}, respectively. We
can see that the self-energy $\Delta \Sigma $ from (\ref{eq42}) is smoother
than in the weak-coupling theory, since the imaginary part of the total
self-energy does not vanish at the Fermi energy. The mean-field self-energy $%
\Sigma -\Delta \Sigma $ is dominant in the vicinity of the Fermi energy,
while $\Delta \Sigma $, due to the ring diagrams, is responsible for the
band tails.

\section{Conclusions}

In this paper we derived a general expansion scheme around
mean-field theories for interacting electrons. We used a quantum version of
the linked cluster expansion modified for the needs of perturbing the
mean-field theories under consideration. We used the difference of the full
and the mean-field Hamiltonian as the perturbation generating the LCE. We
succeeded in this way to put expansions around weak-coupling (Hartree-Fock)
and strong-coupling (Hubbard-III) mean-field theories on the same footing.
Moreover, the proposed expansion scheme applies to the grand potential (free
energy) and hence the self-consistent theory has a variational character and
is thermodynamically consistent and conserving in the Baym sense. The
presented application of the LCE around mean fields enables to
systematically classify, renormalize and sum various classes of diagrams as
is usual in the weak-coupling expansion, although the Wick theorem and the
Gaussian integration are not used.

We explicitly investigated an approximation using a sum of
the so-called ring diagrams leading at the classical level to the
Bloch-Langer approximation \cite{bl65}. We showed that this approximation at
the quantum level, where additive internal degrees of freedom represented by
Matsubara frequencies are to be involved, is not solvable. Further
approximations are enforced to obtain the grand potential in closed form. We
discussed and compared various approximations resulting from simplifications
of the complete, renormalized sum of the ring diagrams. Thereby we paid
special attention to the problem of renormalizations and the
self-consistency of the derived approximate solutions.

At first we studied the influence of dynamical
fluctuations due to the $U^2$ term. When expanded around the Hartree-Fock
solution, we obtained a modified iterated perturbation theory of \cite{gk92}%
. The difference to the IPT was manifested by a complex prefactor in the
self-energy. This prefactor is a consequence of a partial self-consistency
of the approximation and of the fact that the LCE is a diagrammatic
expansion for the grand potential, whereas the IPT for the self-energy.
These prefactors are ubiquitous feature of non-self-consistent or only
partially self-consistent theories, where the Ward identities and sum rules
need not be fulfilled. Although there are no unphysical complex prefactors
in the IPT for the self-energy, the simple analytic structure of e.g.\
susceptibilities goes lost. There is no set of diagrams
generating susceptibilities in the IPT. We found that incompletely
self-consistent theories containing the $U^2$ term break down numerically at
a critical $U_c$ above which no stable numerical solution for complex
energies was found. This situation did not improve even if we took into
account the ring diagrams without any dynamical renormalizations. A pole in
the two-particle function destroys convergence of the iterations.

We concluded from the difficulties with nonrenormalized
partial summations of the LCE that only dynamically renormalized, fully
self-consistent theories can provide global, thermodynamically consistent
approximations. We hence developed a scheme how to introduce full
self-consistency into the LCE in a controlled and systematic way. We
demonstrated the method on an example of the ring diagrams, where we showed
that a class of ``noncrossing diagrams'', when summed, leads to full
self-consistency of the approximation. Such a fully self-consistent theory
has a diagrammatic expansion for any derivative of the free energy. The
self-consistent version of the ring diagrams around the Hartree-Fock
solution was identified with the bubble-chain approximation studied at the
level of the self-energy in \cite{mm91}. We studied approximations with the
self-consistent ring diagrams for both the weak and strong coupling
mean-field theories. In the weak-coupling case we found that for
intermediate interaction strengths a pole in the two-particle function
interferes and hinders to reach convergence of iterations. A more advanced
approach to handle the behavior of the iterations in the vicinity of the
pole must be used to decide how the pole is being approached. This problem,
tightly connected with the metal-insulator transition, will be discussed in
a separate publication. The ring-diagram approximation around the
strong-coupling MFT was found to be stable for all interaction strengths and
to lead to a metal-insulator transition at about $U_c\simeq 5.5$ in the same
way as the static MFT of \cite{vjdv92}.

The linked cluster expansion with dynamical
renormalizations introduced in this paper is not only suitable to derive
conserving approximations with dynamical fluctuations beyond static
mean-field theories, but it may prove efficient for expansions around other
nontrivial solutions such as the expansion for the Kondo model proposed
earlier by Anderson and Yuval \cite{anyu69}.

We thank D.Vollhardt for useful discussions.
The work was supported in part by the grant No.
202/95/0008 of the Grant Agency of the Czech Republic and by the
Sonderforschungsbereich 341 of the Deutsche Forschungsgemeinschaft.

\appendix

\section{Electron-hole symmetry}

We now prove that the Green function $G(z)$ and
consequently the self-energy $\Sigma(z)$ transform under the electron-hole
transformation defined as
\begin{eqnarray}
\label{a1}
z\to -z &\hspace{1cm}& n_\sigma \to 1-n_{-\sigma}
\end{eqnarray}
according to formulas (\ref{eq33}).

We use the results of the static mean-field theory \cite
{vjjmdv93} and the fact that the ring diagrams do not change the equations
for the mean-field parameters $n^s$ and $E$ to obtain the transformation
rules
\begin{eqnarray}
\label{a2}
n^s_\sigma\to 1-n^s_{-\sigma}\;,\hspace{1cm} & \mu\to \frac U2 -\mu\;, &
\hspace{1cm} E\to -E\; .
\end{eqnarray}
Note that here we have to replace the interaction strength $U$ from \cite
{vjjmdv93} with $U/2$ in the transformation formulas, since we use the
convention $\lambda_\sigma=1/2, \lambda_{at}=0$ instead of $%
\lambda_\sigma=1,\lambda_{at}=-1$.

To prove the electron-hole symmetry of the ring-diagram
approximation we assume the validity of (\ref{eq33}) and derive
transformations for the functions $G^{(0)}(z)$ and $G^{(U)}(z)$ and the
integrals $I^{(0)}(z)$ and $I^{(U)}(z)$ and then insert them into the
r.h.s.\ of (\ref{eq29}) to show that the left as well as the right hand side
transform identically.

Using the definitions in eqs.\ (\ref{eq22}), (\ref{eq26a}%
), and (\ref{eq33}) we easily obtain
\begin{eqnarray}
\label{a3}
G^{(0)}(z)\to -G^{(U)}(-z)\;, &\hspace{1cm}& G^{(U)}(z)\to -G^{(0)}(-z)\; .
\end{eqnarray}
Further we must find transformation rules for the integrals $I^{(a)}(z)$. If
we extend the integration in these integrals onto the whole real axis we
obtain
\begin{eqnarray}
\label{a4}
&&\sum_\alpha \frac \alpha{2\pi i} \int_{-\infty}^\infty dx\left\{C_\alpha
(z-x)\left[G^{(a)}(x+i0^+)-G^{(a)}(x-i0^+)\right]\right.\nonumber\\[8pt]
&&\left.+G^{(a)}(z+x)\left[C_\alpha(-x+i0^+)-
C_\alpha(-x-i0^+)\right]\right\}=0\; ,
\end{eqnarray}
which follows from the analyticity of the functions $G^{(a)}(z)$ and $%
C_\alpha(z)$ in the upper and lower complex half-planes. This enables to
deform conveniently the integration contour into the complex plane and split
it into segments containing no singularities.

Using now (\ref{a4}) and the transformation rules (\ref{a3}%
) we obtain
\begin{eqnarray}
\label{a5}
I^{(0)}(z)\to -I^{(U)}(-z)\;, &\hspace{1cm}& I^{(U)}(z)\to -I^{(0)}(-z)\; .
\end{eqnarray}
When we insert (\ref{a3}), (\ref{a5}) into the r.h.s.\ of (\ref{eq29}) we
obtain
\begin{equation}
\label{a6}G(z)\to -G(-z)
\end{equation}
for both sides of (\ref{eq29}) and hence the equations of motion obey the
electron-hole symmetry.

\newpage
\noindent
{\bf Figure Captions}
\medskip

\begin{description}
\item[Fig.1]  DOS at weak coupling calculated from the LCE
with $U^2$-corrections to the Hartree-Fock MFT at $U=1.4$ (broken line) and $%
U=1.7$ (solid line). We used the semi-elliptic DOS with the bandwidth $2D=4$.

\item[Fig.2]  DOS at $U=1.9$ in the same theory as in
Fig.1. For larger $U$'s the iterations do not converge any longer.

\item[Fig.3]  DOS calculated with the iterated
perturbation theory (IPT) within the domain of convergence of iterations, $%
U=1.7,2.7,4.7$ depicted as dotted, broken and solid lines, respectively.

\item[Fig.4a]  Real part of the propagator $G^{(0)}$
within IPT for $U=1.7,3.7,4.7$ as in Fig.3.

\item[Fig.4b]  Imaginary part of $G^{(0)}$ within IPT for
the same values of $U$ as in Fig.4a.

\item[Fig.5]  The location $\omega _{min}$ of the minima
in Re$G^{(0)}(\omega )$ (broken line) and Im$G^{(0)}(\omega )$ (solid line).

\item[Fig.6]  DOS for the IPT extended of the ring
diagrams at $U=1.87$.

\item[Fig.7]  Simplest diagrams contributing to the
expansion of the exponential in (\ref{eq20}). The hatched bubbles represent
the function $C$, the wavy lines derivatives w.r.t. generating sources.

\item[Fig.8]  DOS for the fully self-consistent
approximation with the ring diagrams around the Hartree-Fock MFT at $U=1.5$
(broken line) and $U=3.5$ (solid line).

\item[Fig.9]  DOS for the fully self-consistent
approximation with the ring diagrams around the strong-coupling MFT at $U=1.5$
(broken line), $U=3.5$ (dotted line), and $U=5$ (solid line).

\item[Fig.10a]  Contribution to the imaginary part of the
self-energy $\Delta \Sigma $ due to the ring diagrams at $U=3.5$ for the
strong-coupling (broken line) and the Hartree-Fock (solid line) MFT.

\item[Fig.10b]  Contribution to the imaginary part of the
self-energy $\Sigma -\Delta \Sigma $ due the Hubbard-III term in the static
strong-coupling MFT (broken line) and the dynamic MFT with ring diagrams
(solid line) at $U=3.5$.
\end{description}


\begin{references}
\bibitem{note1}  An approximation is called $\Phi $%
-derivable if the free-energy functional can be written in closed form,
see ref.\ \cite{wor74}.
\end{references}
\end{document}